\documentclass[preprint,showpacs,preprintnumbers,amsmath,amssymb,nofootinbib]{revtex4-1}

\usepackage{graphicx}
\usepackage{dcolumn}
\usepackage{bm}
\usepackage{color}
\usepackage[mathscr]{eucal}

\newcommand{\beq}{\begin{equation}}
\newcommand{\eeq}{\end{equation}}
\newcommand{\beqy}{\begin{eqnarray}}
\newcommand{\eeqy}{\end{eqnarray}}
\newcommand{\beqyn}{\begin{eqnarray*}}
\newcommand{\eeqyn}{\end{eqnarray*}}
\newcommand{\nl}{\newline}
\newcommand{\nn}{\nonumber}

\newcommand{\slas}[1]{\not\!{#1}}

\newcommand{\bc}{\begin{center}}
\newcommand{\ec}{\end{center}}
\newcommand{\bmin}{\begin{minipage}}
\newcommand{\emin}{\end{minipage}}

\usepackage{latexsym}

\usepackage{dsfont}
\usepackage{multirow}

\newcommand{\bea}{\begin{eqnarray}}
\newcommand{\eea}{\end{eqnarray}}
\newcommand{\be}{\begin{equation}}
\newcommand{\ee}{\end{equation}}

\newlength\savedwidth

\newcommand{\lind}[2]{^{\phantom{#1}#2}_{#1}}

\begin{document}

\title{On the role of the subsidiary fields in the momentum and angular momentum in covariantly quantized QED and QCD}

\author{Elliot Leader}
 \email{e.leader@imperial.ac.uk}
\affiliation{Blackett laboratory \\Imperial College London \\ Prince Consort Road\\ London SW7 2AZ, UK}

\date{\today}

\begin{abstract}
 The covariant quantization of QED and QCD requires the introduction of subsidiary gauge-fixing and ghost fields and it is crucial to understand the role of these in the energy-momentum tensor, and in the momentum and angular momentum operators. These issues were studied in  \cite{Leader:2011za}, and key results from this study  were utilized in the major review of angular momentum in \cite{Leader:2013jra}. Damski \cite{Damski:2021feu} has rightly criticized as incorrect the derivation of certain equations in \cite{Leader:2011za}. We show, however,  that the key results of \cite{Leader:2011za} which are utilized in \cite{Leader:2013jra} are unaffected by Damski's criticism.
\end{abstract}

\maketitle

\section{\label{Intro} Introduction}
 It is well known that the covariant quantization of QED and QCD i.e. in which the photon vector potential $A^\mu(x)$ and the gluon vector potential $A^\mu_a$ transform as genuine Lorentz 4-vectors, is a non-trivial task \cite{Lautrup, Nakanishi:66,Nakanishi:72,Kugo:1979gm} involving the introduction of a scalar gauge-fixing field (Gf) in QED and both a gauge-fixing field and Faddeev-Popov ghosts fields (Gf+Gh) in QCD. In both QED and QCD  the expressions for the linear and angular momentum \emph{operators}  include terms involving all these fields.\nl
 As explained in detail in \cite{Leader:2011za} it is  necessary to work in an indefinite-metric space i.e. one in which the ``length" or norm of a vector can be either positive or negative and the definition of the ``physical states" with positive norm has to specified with care. It is also necessary to specify the condition for an operator to represent a physically measurable quantity i.e. to be an ``observable". \nl
 The situation is further complicated by the fact that it is possible to deal with different versions of the energy-momentum tensor $t^{\mu\nu}$, of which the two most important are the Canonical version $t^{\mu\nu}_{can}$ which follows from Noether's theorem and the Bellinfante version $t^{\mu\nu}_{bel}$ which is symmetric under $(\mu\leftrightarrow \nu)$ and which differs from $t^{\mu\nu}_{can}$ by a divergence term, as will be spelled out in detail below. Both these versions of the energy-momentum tensor are conserved quantities. Based on these one can define, in the standard way, the momentum operators $P^\mu_{can}$ and  $P^\mu_{bel}$ and the angular momentum operators $J^i_{can}$ and $J^i_{bel}$. Because the energy-momentum tensors  are related by a divergence one can show that for the matrix elements between any  normalizable physical states, $\langle  \Phi |\, P^\mu_{bel} \, |  \Psi \rangle = \langle  \Phi |\, P^\mu_{can}  \,|  \Psi \rangle $ and $\langle  \Phi |\, J^i_{bel} \, |  \Psi \rangle = \langle  \Phi |\, J^i_{can}  \,|  \Psi \rangle $.\nl
 Of key physical interest in QCD is the question of the fraction of the momentum and angular momentum carried by quarks and gluons in a hadron, with analogous questions about electrons and photons in QED. Clearly to answer these questions one has to know the contributions of the subsidiary fields to the physical matrix elements of the above  operators. In all papers dealing with the QCD case it is \emph{assumed, without comment}, that the contribution from the subsidiary fields is zero. A proof of this was given in \cite{Leader:2011za}, which was based on the demonstration that for physical matrix elements in QED $\langle  \Phi'   |\,  t^{\mu\nu}_{bel}(Gf) \, |  \Phi \rangle =\langle  \Phi'   |\,  t^{\mu\nu}_{can}(Gf) \, |  \Phi \rangle =0$ and for QCD $\langle  \Phi'   |\,  t^{\mu\nu}_{bel}(Gf+Gh) \, |  \Phi \rangle =\langle  \Phi'   |\,  t^{\mu\nu}_{can}(Gf+Gh) \, |  \Phi \rangle =0$.\nl
Damski \cite{Damski:2021feu} has pointed out that the proof for both the Canonical and Bellinfante cases in QED is wrong, because  it assumes that the physical states alone form a complete set and thus uses $1=\sum|\Phi\rangle\langle \Phi|$, which is an incorrect ``resolution of the identity", because it leaves out the states of negative norm. Although he does not comment on QCD, Damski's argument shows that the proof that $\langle  \Phi'   |\,  t^{\mu\nu}_{can}(Gf+Gh) \, |  \Phi \rangle =0$ in \cite{Leader:2011za} is also incorrect.\nl
This criticism  seems to imply that the conclusion reached in \cite{Leader:2011za} that the subsidiary fields do not contribute to the momentum and angular momentum is false, but as will be shown, this implication is wrong. \nl
 We shall show the following:\nl
 a) while, indeed, for the Canonical case
 \beq \label{not01} \langle  \Phi'   |\,  t^{\mu\nu}_{can}(Gf) \, |  \Phi \rangle \neq 0 \quad \textrm{and} \quad \langle  \Phi'   |\,  t^{\mu\nu}_{can}(Gf+Gh) \, |  \Phi \rangle \neq 0 \eeq
 on the contrary, for the Bellinfante case,  as claimed in \cite{Leader:2011za}
 \beq \label{belis0} \langle  \Phi'   |\,  t^{\mu\nu}_{bel}(Gf) \, |  \Phi \rangle = 0 \quad \textrm{and} \quad \langle  \Phi'   |\,  t^{\mu\nu}_{bel}(Gf+Gh) \, |  \Phi \rangle =0. \eeq
 b) despite Eq.~(\ref{not01}) one has
 \beq  \partial_\mu \,\langle  \Phi'   |\,  t^{\mu\nu}_{can}(Gf) \, |  \Phi \rangle = \partial_\mu \,\langle  \Phi'   |\,  t^{\mu\nu}_{can}(Gf+Gh) \, |  \Phi \rangle =0 \eeq
 and obviously an analogous result for the Bellinfante case as a consequence  of (\ref{belis0}),
 which are  key results utilized in the major review \cite{Leader:2013jra} of the "Angular Momentum Controversy".\nl
 c) despite Eq,~(\ref{not01}) it turns out that the subsidiary fields do not contribute to the physical matrix elements of either the Canonical or Bellinfante versions of the momentum or angular momentum. This crucial result, as mentioned above, is normally assumed without comment in papers on QCD.

  \section{Physical matrix elements of the Bellinfante energy-momentum tensor  }

We shall show,  as claimed in \cite{Leader:2011za},  that the physical  matrix elements of the gauge-fixing and ghost
  contributions  $t^{\mu\nu}_{bel}(Gf+Gh)$ in QCD and the gauge-fixing contribution      $t^{\mu\nu}_{bel}(Gf)$    in QED,
vanish.   The proof given for the QED case in \cite{Leader:2011za}, as pointed out by Damski \cite{Damski:2021feu}, is incorrect, but the result is actually true .
  Surprisingly   it turns out that the QCD case is simpler to deal with than the QED case , which can be derived as
a special case.
\subsection{\label{QCD} Quantum Chromodynamics}

The pure quark-gluon Lagrangian ${\cal{L}}_{qG}$ is

 \beq \label{Eq:LqG} {\cal{L}}_{qG}= -\frac{1}{4}G^a_{\mu\nu}G^{\mu\nu}_a + \frac{1}{2}\bar{\psi}^l[\delta_{lm}\,i\,(\overrightarrow{\slas{\partial}} -
\overleftarrow{\slas{\partial}} ) - 2 \,g t^a_{lm}\, \not{\negthickspace}{A}^a] \psi^m .\eeq

In order to quantize the theory covariantly one has to introduce both a gauge-fixing field $B(x)$ and Fadeev-Popov anti-commuting fermionic ghost fields $c(x), \,\bar{c}(x)$.  The Kugo-Ojima Lagrangian \cite{Kugo:78}  for the covariantly quantized theory is then

  \beq \label{Eq:KO} {\cal{L}} = {\cal{L}}_{qG} + {\cal{L}}_{Gf+Gh} \eeq

  where

  \beq \label{Eq:Gf+Gh} {\cal{L}}_{Gf+Gh} =  -
  i(\partial^\mu\bar{c}^a)D^{ab}_\mu c_b  - (\partial^\mu B^a) A^a_\mu + \frac{\textsf{a}}{2}B^aB^a . \eeq

The physical states $|\Psi \rangle $ are defined by the subsidiary conditions
\beq \label{Eq:Phys} Q_B |\Psi \rangle =  0  \eeq
\beq  \label{Eq:Phys'} Q_c |\Psi \rangle  = 0 \eeq

where the conserved, hermitian charge $Q_B$ is given by

\beq \label{Eq:QBdef} Q_B= \int d^3x [ B^a {\overleftrightarrow{\partial}}_0c^a -gB^af_{abc}A^b_0c^c -i(g/2) (\partial_0\bar{c}^a) f_{abc}c^bc^c]. \eeq

and  the conserved charge $Q_c$

\beq \label{Eq:Qc} Q_c=\int d^3x [\bar{c}^a\overleftrightarrow{\partial}_0 c^a -g \bar{c}^af_{abc}A^b_0c^c] \eeq

 ``measures" the \emph{ghost number}

\beq \label{Eq:GNo} i[Q_c, \phi] = N \phi \eeq

where $N=1$  for  $\phi=c,\,\, -1 \,\, \textrm{for}\,\, \phi=\bar{c} \,\, \textrm{and} \,\, 0$   for all other fields.

The Bellinfante energy-momentum tensor is
\beq \label{Eq:tbQCD} t^{\mu\nu}_{bel}=  t^{\mu\nu}_{bel}(qG) + t^{\mu\nu}_{bel}(Gf+Gh) \eeq

where
\beq \label{Eq:tbelqG} t^{\mu\nu}_{bel}(qG)= \frac{i}{4}[\bar{\psi}_l \gamma^\mu \overleftrightarrow{D}^\nu \psi_l + (\mu \leftrightarrow \nu )] - G^{\mu\beta}_a G^\nu_{a\beta}- g^{\mu\nu}{\cal{L}}_{qG}  \eeq

and the gauge-fixing and ghost terms are given by
\beq \label{Eq:tbelGf} t^{\mu\nu}_{bel}(Gf+Gh)= - (A^\mu_a \partial^\nu B_a + A^\nu_a \partial^\mu B_a) -i[(\partial^\mu\bar{c}_a)D^{\nu}_{ab} c_b +(\partial^\nu\bar{c}_a)D^{\mu}_{ab} c_b ] - g^{\mu\nu} {\cal{L}}_{Gf+Gh}. \eeq

This can be rewritten \cite{Kugo:1979gm} as an anti-commutator with $Q_B$
\beq \label{tbelGfnew} t^{\mu\nu}_{bel}(Gf+Gh)= - \{ Q_B, \, \big( (\partial^{\mu} \bar{c}_a) A^{\nu}_a + (\partial^{\nu} \bar{c}_a) A^{\mu}_a + g^{\mu\nu}[\frac{ \texttt{a}}{ 2}\bar{c}_aB_a - (\partial^{\rho}\bar{c}_a) A_\rho^a ] \big) \}. \eeq

It follows from Eqs.~(\ref{Eq:Phys}) and (\ref{tbelGfnew})  that $t^{\mu\nu}_{bel}(Gf+Gh)$  does not contribute to physical matrix elements i.e.
\beq \label{Gf+Gh=0}\langle  \Phi'   |\,  t^{\mu\nu}_{bel}(Gf+Gh) \, |  \Phi \rangle =0 \eeq so that
\beq \label{tbelQCD} \langle  \Phi'   |\,  t^{\mu\nu}_{bel, QCD} \, |  \Phi \rangle = \langle  \Phi'   |\,  t^{\mu\nu}_{bel}(qG) \, |  \Phi \rangle \eeq
and hence that the subsidiary fields do not contribute to the QCD expressions for either $P^\mu_{bel}$ or $\bm{J}_{bel}$ i.e.
\beq  \label{defPmu}\langle  \Phi   |\,  P^\mu_{bel} \, |  \Phi \rangle\equiv\int d^3 x \langle  \Phi   |\,  t^{0\mu}_{bel} \, |  \Phi \rangle = \int d^3 x \langle  \Phi   |\,  t^{0\mu}_{bel}(qG) \, |  \Phi \rangle = \langle  \Phi   |\,  P^\mu_{bel}(qG) \, |  \Phi \rangle \eeq and
\beqy \label{defJi} \langle  \Phi   |\,  J^i_{bel} \, |  \Phi \rangle \equiv \frac{1}{2} && \epsilon^{ijk}\ \int d^3x\, \langle  \Phi   |\, x^j t^{0k}_{bel} \, -   x^k t^{0j}_{bel} \, |  \Phi \rangle \nn \\ =\frac{1}{2} &&\,\epsilon^{ijk}\, \int d^3x\,\langle  \Phi   |\, x^j t^{0k}_{bel}(qG) \,  - x^k t^{0j}_{bel}(qG) \, |  \Phi \rangle = \langle  \Phi   |\,  J^i_{bel}(qG) \, |  \Phi \rangle . \eeqy

\subsection{\label{QED} Quantum Electrodynamics}

The most general covariantly quantized version of QED is given by the Lautrup-Nakanishi Lagrangian density \cite{Lautrup, Nakanishi:66}, which is a combination of the Classical Lagrangian ($Clas$) and a Gauge Fixing part ($Gf$)

 \beq \label{Eq:LN} {\cal{L}} = {\cal{L}}_{Clas}  +  {\cal{L}}_{Gf} \eeq

 where

 \beq \label{Eq:LClas} {\cal{L}}_{Clas} = -\frac{1}{4}F_{\mu\nu}F^{\mu\nu}
  +  \frac{1}{2}\, [\bar{\psi} (i\slas{\partial} - m + e \not\negthickspace{A} )\psi +  \textrm{h.c.}]  \eeq

  and

 \beq \label{Eq:Gf} {\cal{L}}_{Gf} = -\partial_{\mu}B(x).A^{\mu}(x)+ \frac{\textsf{a}}{2}B^2(x) \eeq

  where $B(x)$ is the gauge-fixing field and the parameter $ \textsf{a} $ determines the structure of the photon propagator and is irrelevant for the present discussion\footnote{The case $\textsf{a}=1$ corresponds to the Gupta-Bleuler approach (see e.g. \cite{Jauch:1955jr}) based on the Fermi Lagrangian. Note also that in order to conform to the conventions used in the QCD case, the expression for ${\cal{L}}_{Gf}$ differs from Nakanishi-Lautrup by a 4-divergence.}

  The \emph{physical} states $|\Phi \rangle $ of the theory are defined to satisfy

  \beq \label{Eq:phys} B^{(+)}(x) | \Phi \rangle = 0  \eeq

  where

  \beq \label{Eq:B} B(x)  =B^{(+)}(x) +   B^{(-)}(x)  \eeq

  with $B^{(\pm)}(x)$  the positive/negative frequency parts of $B(x)$.

For the conserved Bellinfante density one finds,

\beq \label{Eq:tbQED}  t^{\mu\nu}_{bel}  = \theta^{\mu\nu}_{bel} + t^{\mu\nu}_{bel}(Gf) \eeq

where $\theta^{\mu\nu}_{bel}$, which is referred to as the classical energy momentum tensor density, is

\beq \label{Eq:tClas} \theta^{\mu\nu}_{bel} =  \frac{i}{4}\, \bar{\psi}(\gamma^\mu \overleftrightarrow{D}^\nu + \gamma^\nu \overleftrightarrow{D}^\mu )\,\psi - F^{\mu\beta}F^\nu_{\phantom{\nu}\beta} - g^{\mu\nu}{\cal{L}}_{Clas} \eeq

where $\overleftrightarrow{D}^\nu = \overleftrightarrow{\partial}^\nu -2ieA^\nu $,
and

\beq \label{Eq:tGf}  t^{\mu\nu}_{bel}(Gf) = - A^\nu\,\partial^\mu B -A^\mu\,\partial^\nu B - g^{\mu\nu}{\cal{L}}_{Gf}  \eeq

As explained in Kugo-Ojima \cite{Kugo:78} the QED expressions  can be obtained from the QCD case by putting the structure constants to zero and suppressing the colour group labels, in whch case the gauge-fixing field $B$ and the ghost fields $c$ and $\bar{c}$ become free fields. The expression Eq.~(\ref{Eq:QBdef})  for the charge $Q_B$  then becomes, in terms of creation and annihilation operators
\beq \label{QBqed} Q_B = i \int\frac{d^3 k}{(2\pi)^3 2 E_k} [c_k^\dag B_k - B_k^\dag c_k]. \eeq
Because the Fadeev-Popov ghosts are here free, the state vector space $\mathcal{V}$ can be decomposed into a direct product
 $\mathcal{V}=\mathcal{V}_{phys} \otimes   \mathcal{V}_{FP}$ where $\mathcal{V}_{phys}$ is the usual QED physical state vector space. Moreover the ghosts are redundant, so that one can work in the sector containing neither $c$ nor $\bar{c}$ ghosts i.e. $\mathcal{V}_{phys} \otimes | 0 \rangle _{FP}$. Hence the physical states in QED can be taken to be $|\Phi\rangle \otimes |0\rangle_{FP} $. Using this and (\ref{QBqed}), Eq.~(\ref{Eq:Phys}) can then be shown to imply (\ref{Eq:phys}). Hence
  \beq \langle  \Phi'   |\,  t^{\mu\nu}_{bel, QED}(Gf) \, |  \Phi \rangle = _{FP}\hspace{-0.1cm}
  \langle 0 |\otimes\langle  \Phi'   |\,  t^{\mu\nu}_{bel,Free}(Gf+Gh) \, |  \Phi \rangle\otimes|0\rangle_{FP} =0  \eeq
  where $t^{\mu\nu}_{bel,Free}(Gf+Gh) $ is $t^{\mu\nu}_{bel,QCD}(Gf+Gh)$ in which the structure constants are put to zero and the gauge-fixing and ghost fields are free. Thus
  \beq \label{tbelQED} \langle  \Phi'   |\,  t^{\mu\nu}_{bel, QED} \, |  \Phi \rangle = \langle  \Phi'   |\,  \theta^{\mu\nu}_{bel} \, |  \Phi \rangle \eeq
and hence, as in QCD, the subsidiary fields do not contribute to the QED expressions for the physical matrix elements of either $P^\mu_{bel}$ or $\bm{J}_{bel}$.

\vspace{2cm}
\subsection{\label{Damski} QED: direct study of subsidiary fields}

The argument showing that the gauge fixing field in QED does not contribute to the physical expectation value of the
Bellinfante version of the energy-momentum tensor and hence does not contribute to the expectation values of the
Bellinfante
versions of the momentum and angular momentum, based on the QCD case, is rather abstract, so we here show that the
 concrete expression for the gauge fixing contribution to the Bellinfante angular momentum,  given in Damski's paper \cite{Damski:2021feu}, which deals only with the free electromagnetic case,
actually
vanishes i.e. we shall show directly that
\beq \label{JbelGf} \langle \Phi | \, \mathbf{J}_{bel}(\textrm{Gf}) \,|  \Phi \rangle =0 \eeq
where, in Damski's notation
\beq J^i_{bel}(\textrm{Gf}) = J^i_{\textrm{div}} + J^i_{\xi}. \eeq
Consider
\beq \label{JbelDam} \langle \Phi | J^i_{bel}(\textrm{Gf}) | \Phi \rangle  =  \langle \Phi | J^i_{\textrm{div}} +
J^i_{\xi} | \Phi \rangle
= \int d^3 z  \epsilon^{imn}\langle \Phi | z^m \mathcal{I}_n(z) - z^n \mathcal{I}_m(z) | {\Phi} \rangle  \eeq
with, following Damski,
\beq \mathcal{I}_n(z) = A_n\partial_j F_{0j} + (\partial\cdot A) \partial_n A_0 . \eeq
Defining
\beq B(z)\equiv - \partial\cdot A \eeq
 and using the fact that in Damski the fields are free, one obtains
 \beq \label{In} \mathcal{I}_n(z) = A_n \partial_0 B -B \partial_n A_0 .\eeq
 We split the fields into their positive and negative frequency parts
 \beq B(z)= B^{(+)} + B^{(-)}  \qquad B^{(-)}= [B^{(+)}]^\dag \quad  \textrm{and} \quad    A_\mu (z) = A_\mu^{(+)}  +
A_\mu^{(-)}
\qquad A_\mu^{(-)}= [A_\mu^{(+)}]^\dag \eeq with
 \beq \label{Amu} A_\mu^{(+)}(z) = \int [d k'] \sum_{\sigma=0}^{\sigma=3} \epsilon_\mu( \textbf{k}', \sigma)c_{
\textbf{k}'\sigma}e^{-ik'\cdot z} \eeq
 where the $ \epsilon_\mu$ are polarization vectors and the $c_{ \textbf{k}'\sigma}$ are  annihilation operators and we
use the  shorthand
 \beq [d k'] \equiv \frac{d^3 k'}{(2\pi)^{3/2} \sqrt{2 \omega_{k'}}}. \eeq
 In Damski's notation one has
 \beq \label{B} B^{(+)}(z) = -i \int [dk]\omega_k L_{\mathbf{k}}\omega e^{-ik\cdot z} \eeq
 where
 \beq L_{\mathbf{k}}= c_{\mathbf{k}3} -c_{\mathbf{k}0} .\eeq
The physical states, as usual,  are defined to satisfy
 \beq B^{(+)}(z) | \Phi \rangle =0 \qquad \langle \Phi | B^{(-)}(z)  =0. \eeq
 Using these and the fact that the commutators $[B^{(+)}, A_\mu^{(+)}] =[B^{(-)}, A_\mu^{(-)}]=0 $, it follows that
\beqy \label{GF} \langle \Phi |  \mathcal{I}_n(z)  | \Phi \rangle = \langle \Phi |    A_n \partial_0 B^{(-)} -B^{(+)}
\partial_n A_0     | \Phi \rangle \nn \\
 = \langle \Phi |    [A^{(+)}_n , \partial_0 B^{(-)}]  - [B^{(+)} , \partial_n A^{(-)}_0 ]    | \Phi \rangle \nn \\
 =\langle \Phi | \Phi \rangle \{ [A^{(+)}_n , \partial_0 B^{(-)}]  - [B^{(+)} , \partial_n A^{(-)}_0 ] \} \eeqy
 the last step following since the commutators are c-numbers.  \nl
The relevant commutators and polarization vectors are
\beq \label{Comms}[L_{\bm{k}},c^\dag_{\bm{k}'0}] = - [c_{\bm{k}0},c^\dag_{\bm{k}'0}]= \delta^3(\bm{k}'-\bm{k} ) \qquad
\quad
[L_{\bm{k}},c^\dag_{\bm{k}'3}] = [c_{\bm{k}3},c^\dag_{\bm{k}'3}]= \delta^3(\bm{k}'-\bm{k} ) \eeq
and
\beq \label{Pols} \epsilon^\mu(\bm{k}, 0 )= (1,\bm{0})\qquad   \qquad \epsilon^\mu(\bm{k}, 3 )= (0, \bm{k}/\omega_k).
\eeq
Substituting Eqs.~(\ref{Amu}, \ref{B}, \ref{Comms}, and \ref{Pols}) into Eq.~(\ref{GF}) , we find, after some labour,
that
\beq      [A^{(+)}_n , \partial_0 B^{(-)}]  - [B^{(+)} , \partial_n A^{(-)}_0 ] =0 \eeq
and thus that  in QED
\beq      \langle \Phi | J^i_{bel, QED}(\textrm{Gf}) | \Phi \rangle =0 \eeq
in agreement with the more general derivation based on the QCD case.

\subsection{Bellinfante summary}

To summarize, despite the incorrect proof for the QED case given in \cite{Leader:2011za} , the physical matrix elements of the gauge-fixing and ghost contributions to the Bellinfante version of the energy-momentum tensor, in both QED and QCD, actually do vanish. Hence the only contributions to the momentum $P^\mu_{bel}$ and angular momentum $\bm{J}_{bel}$ are from photons and electrons in the QED case and from quarks and gluons in  QCD. \nl
This latter property was thus correctly utilized in the review paper \cite{Leader:2013jra}. Also, in writing down the most general structure for the matrix elements of $\langle  \Phi'   |\,  t^{\mu\nu}_{bel}(qG) \, |  \Phi \rangle $ in \cite{Leader:2013jra}, use was made of the claim that
\beq \partial_\mu \,\langle  \Phi'   |\,  t^{\mu\nu}_{bel}(qG) \, |  \Phi \rangle =0. \eeq
This follows because  the total $t^{\mu\nu}_{bel}$ is a conserved operator and, via (\ref{tbelQCD}),
\beq \label{consbel}\partial_\mu \,\langle  \Phi'   |\,  t^{\mu\nu}_{bel}(qG) \, |  \Phi \rangle =\partial_\mu \,\langle  \Phi'   |\,  t^{\mu\nu}_{bel} \, |  \Phi \rangle=0. \eeq

Note also that this justifies the results in several papers in the literature, e.g.
   Ji \cite{Ji:1996ek,Ji:1996nm,Ji:1997pf},  Jaffe and Manohar \cite{Jaffe:1989jz},  Bakker, Leader and Trueman (BLT) \cite{Bakker:2004ib} and Wakamatsu \cite{Wakamatsu:2010qj,Wakamatsu:2010cb}, where the general structure of the physical matrix elements of $t^{\mu\nu}_{bel}(qG)$ (or its QED analogue) is derived under the \emph{unstated assumption} that Eq.~(\ref{consbel}) holds.

  \section{Physical matrix elements of the Canonical energy-momentum tensor  }

  In \cite{Leader:2011za} it  was claimed   that the physical  matrix elements of the gauge-fixing and ghost
  contributions  $t^{\mu\nu}_{can}(Gf+Gh)$ in QCD and the gauge-fixing contribution      $t^{\mu\nu}_{can}(Gf)$    in QED,
vanish.   The result given for the QED case in \cite{Leader:2011za}, as pointed out by Damski \cite{Damski:2021feu}, is wrong because of an incorrect use of the "resolution of the identity" for a space with an  indefinite metric, and, although not discussed by Damski, also the QCD result is wrong for the same reason. \nl
Thus, in contrast to the Bellinfante case and contrary to the claims made  in \cite{Leader:2011za}, in QCD
\beq \label{cannot0} \langle  \Phi'   |\,  t^{\mu\nu}_{can}(Gf+Gh) \, |  \Phi \rangle \neq 0 \eeq
and in QED
\beq \langle  \Phi'   |\,  t^{\mu\nu}_{can}(Gf) \, |  \Phi \rangle \neq 0. \eeq
We shall analyze the consequences of these for the QCD case. Completely analogous arguments hold for the case of QED.\nl
There are two questions which have to be answered.  Given (\ref{cannot0}): \nl
1)  Is the analysis of the general structure of the physical matrix elements of the quark-gluon $t^{\mu\nu}_{can}(qG)$ given in \cite{Leader:2013jra} correct? \nl
2) Do the subsidiary fields contribute to the physical matrix elements of the Canonical momentum $P^\mu_{can}$ and  angular momentum $\bm{J}_{can}$ ?

\subsection{ Structure of the physical matrix elements of the quark-gluon $t^{\mu\nu}_{can}(qG)$}

The Bellinfante and Canonical energy-momentum tensors differ from each other by a divergence term of the following form:
\beq\label{TBel}
t^{\mu\nu}_{can} =  t^{\mu\nu}_{bel} - \partial_\lambda G^{\lambda\mu\nu},
\eeq
where the so-called \emph{superpotential} reads
\beq \label{superpot}
G^{\lambda\mu\nu}=\tfrac{1}{2}\left(M^{\lambda\mu\nu}_\text{spin}+M^{\mu\nu\lambda}_\text{spin}+M^{\nu\mu\lambda}_\text{spin}\right),
\eeq
and, crucially, is antisymmetric w.r.t. its first two indices
\beq \label{Gsym} G^{\lambda\mu\nu}=-G^{\mu\lambda\nu}.\eeq \nl
 The $M_{spin}$ involves a sum over all fields and is given in terms of the Lagrangian by
\beq \label{mspin}
M_\text{spin}^{\mu\nu\rho}(x) = -i \sum_{all fields} \,\frac{\partial \cal{L}}{\partial(\partial_\mu \phi_r)}\, (\Sigma^{\nu\rho})\lind{r}{s}\phi_s(x).
\eeq
where $(\Sigma^{\mu\nu})\lind{r}{s}= -(\Sigma^{\nu\mu})\lind{r}{s}$ is an operator related to the spin of the field. For example, for particles with the most common spins, one has
\begin{align} \hspace{-1cm}
\text{spin-$0$ particle}&  &&\phi(x) &(\Sigma^{\mu\nu})\lind{r}{s}&=0,\\
\text{spin-$1/2$ Dirac particle}& && \psi_r(x) & (\Sigma^{\mu\nu})\lind{r}{s}&= \tfrac{1}{2} \left(\sigma ^{\mu\nu}\right)\lind{r}{s},\label{spinop}\\
\text{spin-$1$ particle} & && A_\alpha(x)  &(\Sigma^{\mu\nu})\lind{\alpha}{\beta}  &= i \left(\delta^\mu_\alpha \,g^{\nu\beta} -\delta^\nu_\alpha \, g^{\mu\beta}\right).\label{spin1op}
\end{align}
Examination of the structure of the Lagrangians in Eqs.~(\ref{Eq:LqG}, \ref{Eq:Gf+Gh}) shows that since ${\cal{L}}_{qG}$ does not contain any gauge-fixing or ghost fields and  ${\cal{L}}_{Gf+Gh}$ does not contain any derivatives of $A^\mu_a$ we may write
\beq M_\text{spin}^{\mu\nu\rho}(x)= M_\text{spin}^{\mu\nu\rho}(x)|_{qG} + M_\text{spin}^{\mu\nu\rho}(x)|_{Gf+Gh} \eeq
and thus in Eq.~(\ref{TBel})
\beq \label{G} G^{\lambda\mu\nu}= G^{\lambda\mu\nu}|_{qG} + G^{\lambda\mu\nu}|_{Gf+Gh} \eeq
so that separately
\beq \label{sepqG} t^{\mu\nu}_{can}(qG) =  t^{\mu\nu}_{bel}(qG) - \partial_\lambda G^{\lambda\mu\nu}|_{qG} \eeq
and
\beq  \label{Gh} t^{\mu\nu}_{can}(Gf+Gh) =  t^{\mu\nu}_{bel}(Gf+Gh) - \partial_\lambda G^{\lambda\mu\nu}|_{Gf+Gh}. \eeq
Hence,
\beqy \partial_\mu \,t^{\mu\nu}_{can}(Gf+Gh) &= &\partial_\mu \,t^{\mu\nu}_{bel}(Gf+Gh)- \partial_\mu\partial_\lambda G^{\lambda\mu\nu}|_{Gf+Gh} \nn \\
&=& \partial_\mu \,t^{\mu\nu}_{bel}(Gf+Gh) \quad \textrm{via} \, \,(\ref{Gsym})\eeqy
and thus for the physical matrix elements, using (\ref{Gf+Gh=0}),
\beq  \partial_\mu \,\langle  \Phi'   |\,  t^{\mu\nu}_{can}(Gf+Gh) \, |  \Phi \rangle = \partial_\mu \,\langle  \Phi'   |\,  t^{\mu\nu}_{bel}(Gf+Gh) \, |  \Phi \rangle =0 .         \eeq
Finally, then, since the total $t^{\mu\nu}_{can}$ is a conserved operator, we obtain the key result
\beq \partial_\mu \,\langle  \Phi'   |\,  t^{\mu\nu}_{can}(qG )\, |  \Phi \rangle = \partial_\mu \,\langle  \Phi'   |\,  t^{\mu\nu}_{can} \, |  \Phi \rangle =0 \eeq
and the analysis of the general structure of the physical matrix elements of the quark-gluon $t^{\mu\nu}_{can}(qG)$ given in \cite{Leader:2013jra}, which relied on this property, is correct.

\subsection{Contribution of the subsidiary fields to the physical matrix elements of the Canonical momentum $P^\mu_{can}$ and  angular momentum $\bm{J}_{can}$ }

From Eqs.~(\ref{defPmu}) and (\ref{sepqG})

\beqy \langle  \Phi   |\,  P^\mu_{bel} \, |  \Phi \rangle &=& \int d^3 x \langle  \Phi   |\,  t^{0\mu}_{bel}(qG) \, |  \Phi \rangle \nn \\
&=& \int d^3 x \langle  \Phi   |\,  t^{0\mu}_{can}(qG) \, |  \Phi \rangle + \int d^3 x \langle  \Phi   |\,  \partial_\lambda G^{\lambda0\nu}|_{qG} \, |  \Phi \rangle.\eeqy
By the antisymmetry property (\ref{Gsym}) the last term is actually a 3-dimensional divergence $\partial_i G^{i0\nu}|_{qG}$, yielding a surface term at infinity, which, as always, is assumed to vanish. Thus
\beq \langle  \Phi   |\,  P^\mu_{bel} \, |  \Phi \rangle=\int d^3 x \langle  \Phi   |\,  t^{0\mu}_{can}(qG) \, |  \Phi \rangle=
 \langle  \Phi   |\,  P^{\mu}_{can}(qG) \, |  \Phi \rangle. \eeq
 But
 \beq \langle  \Phi   |\,  P^\mu_{bel} \, |  \Phi \rangle = \langle  \Phi   |\,  P^\mu_{can} \, |  \Phi \rangle \eeq
 so that, indeed,
 \beq \langle  \Phi   |\,  P^\mu_{can} \, |  \Phi \rangle =\langle  \Phi   |\,  P^{\mu}_{can}(qG) \, |  \Phi \rangle \eeq
 and the subsidiary fields do not contribute to the physical matrix elements of $ P^\mu_{can}$.
 A similar argument shows that
 \beq \langle  \Phi   |\,  J^i_{can} \, |  \Phi \rangle =\langle  \Phi   |\,  J^i_{can}(qG) \, |  \Phi \rangle \eeq
 and the subsidiary fields also do not contribute to the physical matrix elements of $J^i_{can}$.

 \subsection{Canonical summary}

To summarize, despite the incorrect proof given in \cite{Leader:2011za} concerning the physical matrix elements of the gauge-fixing contributions in QED and the gauge-fixing and  ghost contributions in QCD to the Canonical version of the energy-momentum tensor,   the essential property used in writing down the most general structure for the QCD matrix elements  $\langle  \Phi'   |\,  t^{\mu\nu}_{can}(qG) \, |  \Phi \rangle $  and the QED matrix elements $\langle  \Phi'   |\,  \Theta^{\mu\nu}_{can} \, |  \Phi \rangle $ in  \cite{Leader:2013jra}, namely, that
\beq \partial_\mu \,\langle  \Phi'   |\,  t^{\mu\nu}_{can}(qG) \, |  \Phi \rangle =0 \quad \textrm{and} \quad \partial_\mu \,\langle  \Phi'   |\,  \Theta^{\mu\nu}_{can} \, |  \Phi \rangle =0\eeq
 is correct.\nl
Moreover, despite the incorrect derivation  in \cite{Leader:2011za}, the only contributions to the physical matrix elements of the momentum $P^\mu_{can}$ and angular momentum $\bm{J}_{can}$ are from photons and electrons in the QED case and from quarks and gluons in  QCD. This latter property was thus correctly utilized in the review paper \cite{Leader:2013jra}. 

\section{Conclusions}
 Damski's criticism \cite{Damski:2021feu} of the \emph{proof} given in \cite{Leader:2011za} for certain properties of the physical matrix elements of the energy-momentum tensor $t^{\mu\nu}$ in QED is valid, and it is also applicable to QCD, so that, contrary to the assertions in \cite{Leader:2011za}, for the Canonical case
 \beq \label{not0} \langle  \Phi'   |\,  t^{\mu\nu}_{can}(Gf) \, |  \Phi \rangle \neq 0 \quad \emph{and} \quad \langle  \Phi'   |\,  t^{\mu\nu}_{can}(Gf+Gh) \, |  \Phi \rangle \neq 0. \eeq
Nonetheless the following crucial features for QED and QCD, utilized in the angular momentum review \cite{Leader:2013jra}, are in fact correct:\nl
a)  for the Bellinfante case, as claimed in \cite{Leader:2011za}
 \beq \label{Belis0} \langle  \Phi'   |\,  t^{\mu\nu}_{bel}(Gf) \, |  \Phi \rangle = 0 \quad \emph{and} \quad \langle  \Phi'   |\,  t^{\mu\nu}_{bel}(Gf+Gh) \, |  \Phi \rangle =0. \eeq
 b) despite Eq.~(\ref{not0}) one has for the Canonical case,
 \beq  \partial_\mu \,\langle  \Phi'   |\,  t^{\mu\nu}_{can}(Gf) \, |  \Phi \rangle = 0 \quad \emph{and} \quad \partial_\mu \,\langle  \Phi'   |\,  t^{\mu\nu}_{can}(Gf+Gh) \, |  \Phi \rangle =0. \eeq
 and obviously an analogous result for the Bellinfante case as a consequence  of (\ref{Belis0}). \nl
c) despite Eq.~(\ref{not0})  the subsidiary fields do not contribute to the physical matrix elements of either the Canonical or Bellinfante versions of the momentum or angular momentum, a result which is normally \emph{assumed without comment} in papers on QCD.



\bibliography{Elliot_General}

\end{document}